# All-electrical valley filtering in graphene systems (I): A path to integrated electro-valleytronics

Feng-Wu Chen[2], Nin-Yuan Lue[1],
Mei-Yin Chou[3,4,5], and Yu-Shu G. Wu[1,2],*

[1] *Department of Electrical Engineering, National Tsing-Hua University, Hsin-Chu 30013, Taiwan, ROC*
[2] *Department of Physics, National Tsing-Hua University, Hsin-Chu 30013, Taiwan, ROC*
[3] *Institute of Atomic and Molecular Sciences, Academia Sinica, Taipei 10617, Taiwan, ROC*
[4] *Department of Physics, National Taiwan University, Taipei 10617, Taiwan, ROC*
[5] *School of Physics, Georgia Institute of Technology, Atlanta, GA 30327, USA*

Probing and controlling the valley degree of freedom in graphene systems by transport measurements has been a major challenge to fully exploit the unique properties of this two-dimensional material. In this theoretical work, we show that this goal can be achieved by a quantum-wire geometry made of gapped graphene that acts as a valley filter with the following favorable features: i) all electrical gate control, ii) electrically switchable valley polarity, iii) robustness against configuration fluctuation, and iv) potential for room temperature operation. This valley filtering is accomplished by a combination of gap opening in either bilayer graphene with a vertical electrical field or single layer graphene on h-BN, valley splitting with a horizontal electric field, and intervalley mixing by defect scattering. In addition to functioning as a building block for valleytronics, the proposed configuration makes it possible to convert signals between electrical and valleytronic forms, thus allowing for the integration of electronic and valleytronic components for the realization of electro-valleytronics.

* Corresponding author. Email: yswu@ee.nthu.edu.tw

## I. INTRODUCTION

It has been established recently that, in addition to charge and spin, electrons in solids also have a valley pseudospin degree of freedom that can be exploited to create valleytronic devices. [1–3] This property is derived from the existence of multiple minima (maxima) of conduction (valence) bands in momentum space. Promising systems to explore the valley-related properties are two-dimensional honeycomb lattices including gapped graphene systems [4–13] and transition metal dichalcogenides [14–21].

Central to valleytronic applications are the devices that generate valley-polarized electrons for valleytronic signal processing. The first experimental demonstration of valley polarization was achieved by optical pumping with circularly polarized light in single layer $MoS_2$ where the valley-contrasting selection rules are valid for optical transitions in the K and K' valleys [15,19–21]. On the other hand, possible electrical valley filters in graphene systems have been proposed or investigated with filtering mechanisms based on, for example, the valley dependence of quantized states in a zigzag graphene nanoribbon [1], the valley Hall effect [2,22–25], the valley-dependent electron scattering at a line defect [26–28], breaking the valley symmetry by employing strain and magnetic field simultaneously [29–33] or an AC external field [34], and utilizing the energy band warping [35,36]. Diffusive valley currents generated with the inverse valley Hall effect have recently been observed in both single layer [22] and bilayer graphene [23,24]. However, experimental implementation of ballistic valley filters is still quite challenging for various reasons: some of the proposed configurations could not be easily fabricated in the laboratory, may not have switchable polarity, or require the presence of magnetic fields. Hence it leaves plenty of room for further improvement in order to make valleytronics a foreseeable reality.

From the application point of view, it is highly desirable to integrate valleytronic components with conventional electronic components, e.g., integrated electro-valleytronics. This approach allows for the application to take full advantage of each component's unique performance by, for example, using electronic devices for storage units to circumvent the valley decoherence problem in valleytronic devices, and using valleytronic devices for processing units to mitigate the power consumption problem in downscaling integrated circuits [37,38]. In this work, a graphene structure suitable for integrated electro-valleytronics is proposed, which performs the function of valley filtering with the following favorable characteristics: i) all electrical gate control, ii) electrically switchable valley polarity, iii) robust performance against configuration fluctuation, and iv) potential for room temperature operation. Based on these features, the proposed structure acts as a building block of valleytronics and allows for the natural integration of both electronic and valleytronic components, with possible conversion between electrical and valleytronic signals, as shown in **Figure 1**.

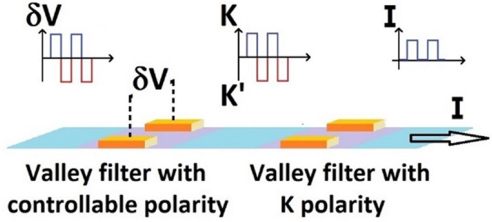

**Figure 1**. Two-way signal conversion. For electrical to valleytronic signal conversion, the time-varying voltage signal (δV) sent into the proposed valley filter produces a time-varying in-plane electric field transverse to the channel of the filter, and the filter outputs a time-varying valley polarized current with polarity varying in phase with the direction of the field. For valleytronic to electrical signal conversion, the valley filter is placed in a fixed polarity (e.g., K) mode. The time-varying valleytronic signal sent into the valley filter results in a time-varying output current (I). The magnitude of the output depends on the polarity of the input current and the polarity set by the filter - a high output if the two are the same and a low output if they are opposite. The output current signal can further be converted into a voltage signal if necessary.

The novel mechanism exploited is the so-called valley-orbit interaction (VOI) that couples the valley pseudospin to an in-plane electric field. The VOI exists in gapped graphene and is similar to the Rashba spin-orbit interaction (SOI), but with a significant difference. As derived previously [37–39], the VOI in single layer graphene (SLG) is given by

$$H_{VOI}^{(MLG)} = \tau \frac{\hbar}{4m\Delta}(\nabla V \times \vec{p}) \cdot \hat{\mathbf{z}}, \qquad (1)$$

which is valley dependent and valley conserving ($\tau = \pm$ being the valley pseudospin index for K/K', $2\Delta$ = energy gap, $m$ = electron effective mass, V = potential energy, $\vec{p}$ = momentum operator, $\hat{\mathbf{z}}$ = unit vector normal to the graphene plane). Some idea about the magnitude of VOI is given below by comparing it with that of SOI in atoms. With $m$ given by $\Delta / v_f^2$ ($v_f$ (Fermi velocity) ~ $10^6$ m/sec) [7], for $\Delta = 0.1$ eV the coefficient "$1/m\Delta$" in VOI is about nine orders of magnitude larger than the corresponding coefficient "$1/m_e^2 c^2$" in SOI ($m_e$ = vacuum electron mass, c = light speed). However, the actual strength of VOI depends on "$\nabla V \times \vec{p}$" used in applications. If we take $\nabla V$ ~ 0.01 meV/Å and p ~ 0.01 $\hbar$ / Å in graphene, and $\nabla V$ ~ 10 eV/Å and p ~ $\hbar$ / Å in atoms, it gives us a VOI strength comparable to or of an order of magnitude larger than the SOI strength. A similar VOI but with a more complex expression also exists in AB-stacking bilayer graphene (BLG) and is given by [37,38]

$$H_{VOI}^{(BLG)} = \frac{1}{4\Delta^2} \Omega_\tau \left\{ H_{21}[V, H_{12}]_- \right\}, \qquad (2)$$

where $H_{12}$ and $H_{21}$ are Hamiltonian matrix elements in the reduced two-band tight-binding model of BLG [8,9], and $\Omega_\tau$ denotes the operation upon an expression by retaining only the τ-dependent terms in the expression.

The device envisioned in this work is basically a quantum wire structure aligned in the armchair direction in gapped graphene, with scattering defects positioned in the vicinity of the wire. The wire constitutes a quasi-one-dimensional current channel and is controlled by electrical gates near the channel. There are various ways to implement the proposed structure in modern laboratories as follows. First, we consider DC-biased AB-stacked BLG for the implementation. It exhibits the following desired properties: i) energy gap tunability with a DC bias between the two layers [8–10], allowing for the patterning of a confined structure [40,41] - a quantum wire in our case, by electrical gating technology; and ii) feasibility to make conventional FETs [42,43] – building blocks for the electronic circuit part of electro-valleytronics. On the other hand, SLG on h-BN can also be used to implement the proposed structure. For example, the quantum wire here can be realized by placing SLG on a trenched h-BN substrate. This gives us a gapless region in the part of graphene suspended above the trench and, simultaneously, creates gapped regions in the substrate-supported part of graphene due to the graphene-substrate interaction [12,13] on both sides of the gapless region. Gapped regions serve as potential barriers and confine carriers in the gapless region resulting in the formation of a quantum wire. Last, we note that BN-doped SLG, e.g., $(BN)_xC_{(1-x)}$ can also be used to form a quantum wire. By varying x, one can create wide- and narrow- gapped regions simultaneously in graphene [44] and confine carriers in the narrow-gapped region.

In this VOI-based proposal, the desirable function of valley filtering is made possible by two key elements. First, electrical gates in the device produce an in-plane electric field transverse to the wire and a corresponding VOI. Due to the valley dependence of VOI, it generates a Rashba-type [45] valley splitting in energy subbands of the wire, as shown in **Figure 2(a)**. Second, scattering defects near the wire induce intervalley K ↔ K' scattering, which couples subband states of opposite valley pseudospins and opens a pseudogap at the crossing point of valley-split subbands, as shown in **Figure 2(b)**. Combining these two features, effective valley filtering can be created as follows. An electronic energy level inside the pseudogap intersects the subbands at two points, corresponding to left- and right-moving electron states, respectively, with these two states carrying primarily opposite pseudospins. A valley polarized current can thus be generated by placing the Fermi level in the pseudogap and applying a small bias to induce electron transport in the wire. Moreover, when the transverse electric field is reversed, valley indices of the two subbands are switched, thus changing the polarity of polarization. In other words, the control of valley polarity can be accomplished by electrical signals, making it possible to construct a functional electro-valleytronic circuit. As an example of applications, we note that two of the proposed valley filters can be put together back to back to form a valley valve – a switching device, where ON/OFF is effected by electrically controlling the relative orientation between valley polarities in the two filters, yielding a low (high) current when the two polarities are opposite (the same).

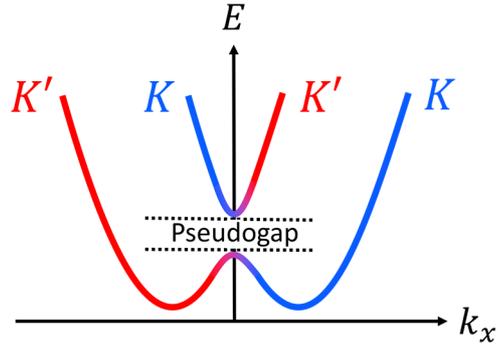

**Figure 2.** Schematic plots. **(a)** Rashba valley splitting in energy subbands of a graphene quantum wire (in the x-direction) with an in-plane electric field (in the y-direction), when intervalley scattering is excluded. **(b)** Opening of a pseudogap when intervalley scattering is included.

Since the proposed filter relies only on the existence of a pseudogap window, it is robust against configuration fluctuation along the channel, as long as the fluctuation does not close the window. We note that an analogous device for spin-filtering has been proposed by Streda and Seba [46]– a semiconductor-based quantum wire filter based on the spin-orbit interaction, where the required pseudogap is opened by an in-plane magnetic field along the wire. The major difference between the two cases lies in the total absence of any magnetic field in our proposal. This can be attributed to the essential distinction between a valley pseudospin and a real spin, which is exploited here to make an all-electrically driven valley filter or valve possible.

This work belongs to a series of our recent studies in ballistic valley filters and valves, which begin with the present proposal of VOI-based valley filters (denoted as **W-I**) and next, use VOI-based valley valves formed of filters proposed here as model systems to investigate various practical effects in generic valley valves as well as VOI-based valley valves (denoted as **W-II**) [47].

This paper is organized as follows. In **Sec. II**, we present the proposed structure and discuss types of the band edge profile transverse to the quantum wire channel. In **Sec. III**, we present the theoretical method for calculation of valley polarization in the proposed filter. In **Sec. IV**,

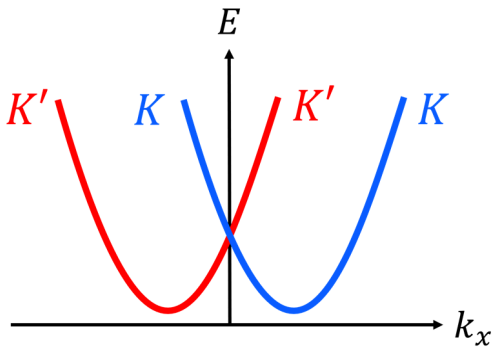

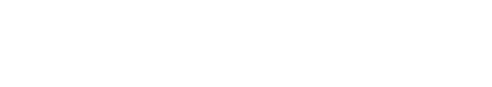

we discuss results of the calculation. In **Sec. V**, we summarize the study.

## II. STRUCTURES AND TRANSVERSE BAND EDGE PROFILES

When implementing the proposed structure, flexible band edge profiles transverse to the channel can be created resulting in different types of quantum wire channels, namely, those of Type-I with straddling band alignment and those of Type-II with staggered band alignment. As shall be discussed below, such a flexibility arises when using BLG for the implementation.

### BLG-based Structures

**Figure 3** shows an example of the proposed valley filter in BLG placed on an insulator substrate. The quantum wire is defined and controlled by both the back gate and the two top gates. In order to achieve valley filtering, it requires the presence of either one or two lines of scattering defects in parallel to the channel. Flexibility in the structure is allowed. For example, the defect lines can be implanted, or replaced by boundaries of graphene oxide regions next to the top gates with the oxide working as scattering defects; the structure can be an armchair nanoribbon with ribbon edges being scattering defects; and the back gate can be single or split. While the former is feasible in practice, from the theoretical perspective, we pick the latter as the structure for study, because it neither requires nor depends on any specific model of scattering defects. In all cases the structure envisioned is consistent with planar processing.

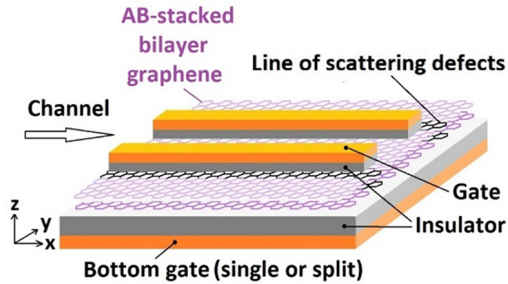

**Figure 3.** The structure of a proposed valley filter in BLG, where BLG is placed on a substrate and is implanted with scattering defects in parallel to the quantum wire channel. Two top gates are placed above BLG, and a (single or split) back gate is placed below the insulator substrate.

**Figure 4** presents the top view of a valley filter in gapped graphene showing various parameters essential to the theoretical calculation. The quantum wire channel is taken to be along the armchair (x) direction. The device is divided into three regions, with $W_2/2 < y < W_1 + W_2/2$ for Region I (under one top gate), $-W_2/2 < y < W_2/2$ for Region II (the channel), and $-W_2/2 - W_3 < y < -W_2/2$ for Region III (under the other top gate). The gap parameter $\Delta$ and the electric potential energy $V$ are taken to be constant in Regions I and III, and a constant and a linear function, respectively, in Region II. The locations of defect lines are given by $y = W_1 + W_2/2$ and $y = -W_2/2 - W_3$.

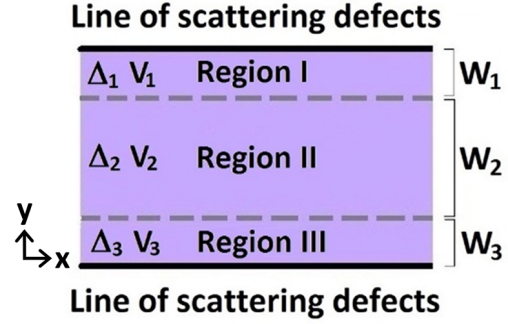

**Figure 4.** Top view of a valley filter, with Region II being the quantum wire channel, and Regions I and III being potential barriers.

In the case of a BLG-based filter, electrical gates are used in the following way. First, DC biases on the back and top gates together create potential differences between the two layers of BLG to open energy gaps in BLG Regions I and III, in order to confine carriers in the channel. Second, the bias difference between top gates creates in BLG an in-plane electric field transverse to the channel to provide the required VOI for valley filtering. Third, the back gate can be either single or split to facilitate the generation of different channel types, as discussed below.

### Transverse band edge profiles

**(Type-I structure)** First, we consider the case of a structure with a Type-I channel, realized with split top gates, split back gates, and a BLG layer placed nearly halfway between the top and back gates, as shown in **Figure 5(a)** along with the distribution of equipotential surfaces. The numerical indices "1", "2", and "3" denote Regions I, II, and III, respectively. Graphene is insulated from the gates by dielectric layers (not shown). For simplicity, the diagram shows the trivial situation where $V_{1t} = V_{3t}$ and $V_{1b} = V_{3b}$, i.e., there is no in-plane potential difference in graphene between Regions I and III. Barrier (channel) graphene is located in regions I/III (II) of a high (low) electric field line density and sees a larger (lower) potential energy drop between the two layers. Denote $E_c$ = conduction band edge, $E_v$ = valence band edge, and $E_g$ = energy gap. **Figure 5(b)** shows the corresponding band edge profile derived from the potential energy distribution in **Figure 5(a)**. As shown in the graph, with the energy gap in BLG determined by the corresponding potential energy drop, it gives a Type-I band alignment across Regions I, II, and III, with $E_g(1) = E_g(3) > E_g(2)$. For confirmation, a numerical simulation of electrical potential energy distribution has been performed for the structure shown in **Figure 5(c)** using the software "Atlas", a semiconductor device simulator. The resultant band edge profile is shown as solid curves in **Figure 5(d)**. Dashed curves are the approximation to solid curves and provide the gap and band offset parameters used in the calculation of **Figure 8** below.

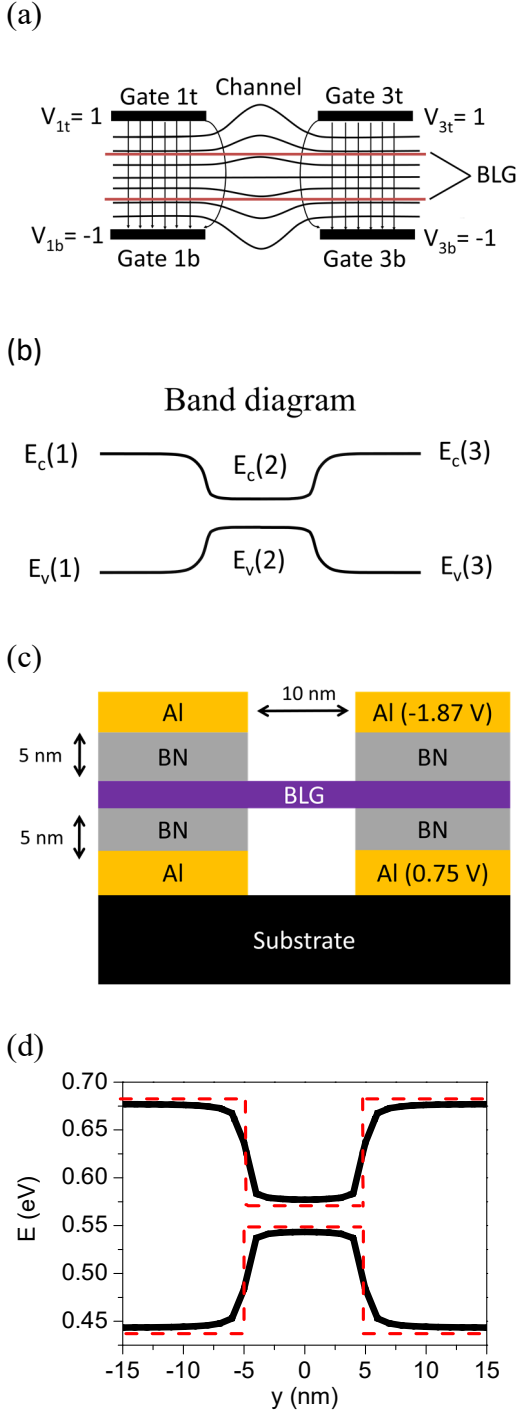

**Figure 5.** (a) Cross section of the BLG-based structure with split bottom gates. Black bars represent the metal gates. Graphene is insulated from the gates by dielectric layers (not shown). $V_{1t(3t)}$ = potential energy at top gate 1t (3t); $V_{1b(3b)}$ = potential energy at back gate 1b (3b). The potential energy is given in arbitrary units. "Arrows" denote electric field lines, and light lines denote the equipotential surfaces. BLG is represented by thick red lines. Numerical indices "1", "2", and "3" denote Regions I, II, and III, respectively. (b) Corresponding band edge profile. (c) Cross section of the structure for electrical potential energy simulation with Atlas. We take $V_{1t} = V_{3t} = 1.87$ eV and $V_{1b} = V_{3b} = -0.75$ eV. (d) Solid curves – conduction and valence band edges obtained from the simulation with the out-of-plane and in-plane dielectric constants of BLG taken to be 1.8 and 3 [48], respectively, and the out-of-plane and in-plane dielectric constants of h-BN taken to be 5.06 and 6.85 [49], respectively. Dashed curves - the approximation to solid curves.

Now if we raise the values of $V_{3t}$ and $V_{3b}$ (potential energies at top and bottom gates in region III, respectively) by the same amount relative to $V_{1t}$ and $V_{1b}$ ((potential energies on top and bottom gates in region I, respectively), the in-plane potential difference at BLG layers, $V_3 - V_1$, grows linearly with $V_{3t} - V_{1t}$, while the gaps in Regions I and III stays the same ($\Delta_1 = \Delta_3$). A reasonable model for the in-plane potential at BLG across the channel would be a linear function, e.g., $\vec{V}(\vec{r}) = V_1 + (V_1 - V_3)(y/W_2 - 1/2)$. This function is to be employed in the calculation of **Figure 8**.

**(Type-II structure)** Next, we consider the configuration of a structure with a Type-II channel, realized with split top gates, a single back gate, and a BLG layer placed approximately midway between the top and back gates, as shown in **Figures 6(a)** along with the distribution of equipotential surfaces.

For simplicity we first consider the situation where $V_{1t} = V_{3t}$. A rough electrostatic potential model is given below. Let $d$ be the BLG thickness and $D$ be the distance between top and back gates. If we assume that the dielectric constants are of the same order for BLG and for the dielectric layers, then, $E_g(1) = E_g(3) \approx V_{1t} \, d/D$. On the other hand, $E_g(2)$ is determined by various factors, e.g., the fringe field and the geometry, e.g., the aspect ratio such as D/(distance between Gate 1t and Gate 3t). Generally, since Region II is immersed in an electric field relatively low in comparison to those in Regions I and III, we have $E_g(1) = E_g(3) > E_g(2)$, as shown in **Figure 6(b)**. Again, for confirmation, a numerical simulation of potential energy distribution has been performed for the structure shown in **Figure 6(c)** using Atlas. The resultant band edge profile is shown as solid curves in **Figure 6(d)**. Dashed curves are the approximation to solid curves and provide the gap and band offset parameters used in the calculation of **Figure 9** below.

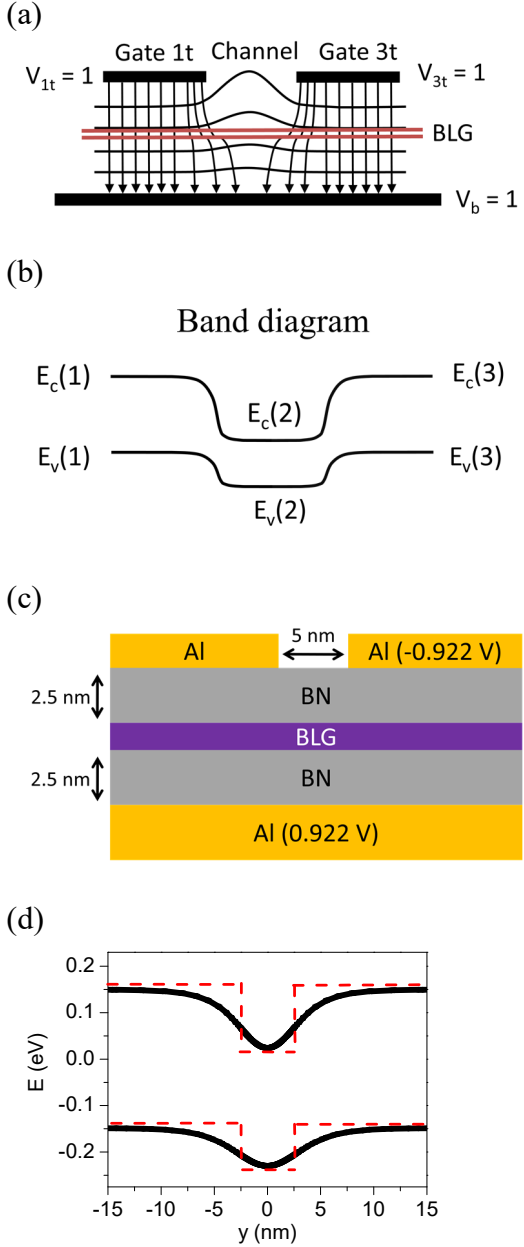

**Figure 6.** (a) Cross section of the BLG-based structure with a single bottom gate. Graphene is insulated from the gates by dielectric layers (not shown). Barrier (channel) graphene is located in Regions I/III (II) of a high (low) electric field line density and sees a relatively high (low) potential energy drop between the two layers. $V_{1t(3t)}$ = potential energy at top gate 1t (3t); $V_b$ = potential energy at back gate. (b) Corresponding band edge profile. (c) Cross section of the structure for potential energy simulation with $V_{1t} = V_{3t} = 0.922$ eV and $V_b$ = -0.922 eV. (d) Solid curves – conduction and valence band edges obtained from the simulation, with the out-of-plane and in-plane dielectric constants of BLG taken to be 1.8 and 3 [48], respectively, and the out-of-plane and in-plane dielectric constants of h-BN taken to be 5.06 and 6.85 [49], respectively. Dashed curves - the approximation to solid curves.

Now, we consider the electrostatic model in a structure with $d/D \ll 1$ in the presence of an in-plane electric field, namely, $V_1 - V_3 \neq 0$ at the BLG layers. It gives $\Delta_1 - \Delta_3 = \frac{1}{2} [E_g(1) - E_g(3)] \sim \frac{1}{2} (V_{1t} - V_{3t}) \, d/D$. Generally speaking, $V_{1t} - V_{3t}$ is of the same order as $V_1 - V_3$. With $d/D \ll 1$, we have $\Delta_1 - \Delta_3 \ll V_1 - V_3$. If $V_1 - V_3$ is additionally set to be of the same order as $\Delta_1$ (or $\Delta_3$), we then have $\Delta_1 - \Delta_3 \ll \Delta_1$ (or $\Delta_3$). Therefore, we conclude that $\Delta_1 \approx \Delta_3$ is a good approximation to the zeroth order of $d/D$ under the condition $d/D \ll 1$, even though $V_1 - V_3$ is of the same order as $\Delta_1$ (or $\Delta_3$). This foregoing approximation is to be used in the calculation of **Figure 9**. For the in-plane potential at BLG across the channel, we again take it to be given by a linear function, e.g., $\vec{V}(\vec{r}) = V_1 + (V_1 - V_3)(y/W_2 - 1/2)$.

We would like to stress that since good filtering capacity shall be demonstrated in the subsequent calculations for both Type- I and II structures, the principle of valley filtering proposed here is expected to be insensitive to the details of the electrostatic model. In other words, we expect that assumptions involved in the electrostatic model can be relaxed or adjusted according to the realistic situation without qualitatively changing the theoretical conclusion.

**SLG-based Structures**

In the case of a SLG-based valley filtering structure, graphene band structures in and outside the quantum wire depend on the SLG-substrate interaction and therefore the band edge profile is fixed without any available flexibility. **Figure 7(a)** presents an example of the proposed structure in SLG/h-BN where SLG is placed on a trenched h-BN substrate. Valley-dependent transport in the quantum wire channel above the trench is controlled by top and back gates. In analogy to the BLG-based structure, it requires either one or two lines of scattering defects, implanted or in the form of oxidation regions or armchair nanoribbon edges. **Figure 7(b)** presents the corresponding transverse band edge profile showing a Type-I channel, with the barrier graphene gap being given by ~ 30 meV [13,22]. Fermi energies in the channel and barriers are taken to be aligned. In the undoped case, due to electron-hole symmetry, the Fermi energy in each region is located midway in the local energy gap thus giving the conduction band offset as ~ 15 meV. The parameters provided here are to be used in the calculation of **Figure 10**.

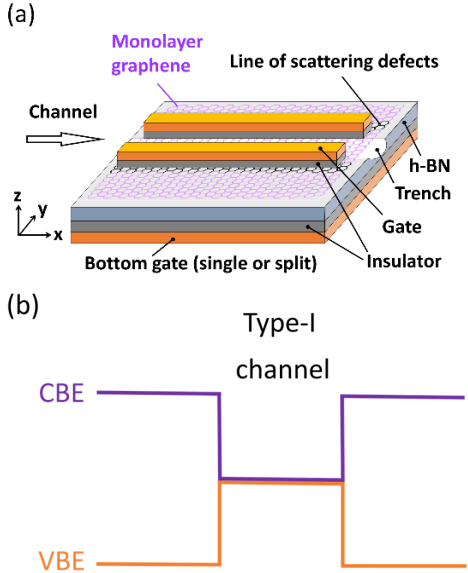

**Figure 7. (a)** The proposed structure of a valley filter in SLG/h-BN, where SLG is placed on a trenched h-BN substrate and is implanted with scattering defects in parallel to the quantum wire channel. Two top gates are placed above SLG, and a (single or split) back gate is placed below the insulator substrate. **(b)** Corresponding transverse band edge profile showing a Type-I channel. CBE: conduction band edge, and VBE: valence band edge.

## III. THEORETICAL METHOD

We study electron transport through the valley filter with the following method. To the lowest-order approximation, we take the channel length to be infinite, calculate the electron energy subband structure of the channel, and infer the electron transport in the channel based on the subband structure. We employ the four-component tight-binding theory of BLG [8–10] for the subband structure calculation, as briefly explained below. Note that when applying it to SLG-based structures, we simply turn off the various interlayer couplings in the theory. The Hamiltonian equation is given by

$$H\Psi_\tau = E\Psi_\tau$$
$$\Psi_\tau(\vec{r}) = \left( \varphi_{A1}^{(\tau)}(\vec{r}) \quad \varphi_{B1}^{(\tau)}(\vec{r}) \quad \varphi_{A2}^{(\tau)}(\vec{r}) \quad \varphi_{B2}^{(\tau)}(\vec{r}) \right)^t,$$
(3)

where H is the Hamiltonian, and $\Psi_\tau$ is the envelope wave function. Here, the subscript or superscript $\tau = +(-)$ denotes the solution near the K (K') valley, with the corresponding wave vector given by $\vec{K} = (0, -4\pi/3\sqrt{3}a)$ ($\vec{K}' = -\vec{K} = (0, 4\pi/3\sqrt{3}a)$) for the K (K') point, and $a = 1.42$ Å is the inter-carbon distance. The four components in $\Psi_\pm$ refer to the wave amplitudes on the four atoms (A1, B1, A2, and B2) in a BLG unit cell. The Hamiltonian H for BLG is given by

$$H = H_0(k_x) + \eta H_1(k_x),$$

$$H_0 = \begin{pmatrix} -\Delta+V & -\gamma_0 & 0 & -\gamma_3 e^{-i3k_x a} \\ -\gamma_0 & -\Delta+V & \gamma_1 & 0 \\ 0 & \gamma_1 & \Delta+V & -\gamma_0 \\ -\gamma_3 e^{i3k_x a} & 0 & -\gamma_0 & \Delta+V \end{pmatrix},$$

$$H_1 = \begin{pmatrix} 0 & -\gamma_0 e^{-i3k_x a/2} & 0 & -\gamma_3 e^{-i3k_x a/2} \\ -\gamma_0 e^{i3k_x a/2} & 0 & 0 & 0 \\ 0 & 0 & 0 & -\gamma_0 e^{-i3k_x a/2} \\ -\gamma_3 e^{i3k_x a/2} & 0 & -\gamma_0 e^{i3k_x a/2} & 0 \end{pmatrix}$$

$$\eta = (\lambda+\lambda^{-1}), \quad \lambda = \exp(\frac{\sqrt{3}}{2} a \partial_y).$$
(4)

Here, $k_x$ is the wave vector along the channel, $\gamma_0$, $\gamma_1$, and $\gamma_3$ are the tight-binding parameters representing various hopping energies, $2\Delta$ is the chemical potential difference between the layers due to the gate biases applied, and V is the potential energy due to the transverse in-plane electric field.

V and $\Delta$ in the Hamiltonian are determined by the various gate biases and are modeled with linear and piecewise constant functions described below. For the three regions depicted in **Figure 4**, we take

Region I ($W_2/2 < y < W_1 + W_2/2$):
$\Delta(\vec{r}) = \Delta_1$, $V(\vec{r}) = V_1$.
Region II ($-W_2/2 < y < W_2/2$):
$\Delta(\vec{r}) = \Delta_2$, $V(\vec{r}) = V_1 + (V_1 - V_3)(y/W_2 - 1/2)$.
Region III ($-W_2/2 - W_3 < y < -W_2/2$):
$\Delta(\vec{r}) = \Delta_3$, $V(\vec{r}) = V_3$.

When an in-plane electric field transverse to the wire is present, we have $V_1 \neq V_3$, and the corresponding potential energy drop in Region II is modeled by a linear function as described above.

We now solve Eqn. (3) for a given $k_x$. Due to time reversal symmetry, the solutions to Eqn. (3) for the two valley pseudospins are degenerate. Therefore, the total wave function is generally of the form

$$\Psi(\vec{r}) = e^{i\vec{K}\cdot\vec{r}}\Psi_+(\vec{r}) + e^{-i\vec{K}\cdot\vec{r}}\Psi_-(\vec{r}) \quad (5)$$

a mixed state of the two degenerate solutions. In view of $\Delta$ being piecewise constant, we proceed as follows.

We divide Region II into M sub-regions, and approximate the linear potential in Region II by a multi-step potential that converges to the linear one in the limit of infinitesimal steps. In our calculation, M = 6 is found to give a reasonably convergent numerical result. Next, we solve for $\Psi(\vec{r})$ in each region of constant potential (labeled by N, N = I, II-1, II-2, …, II-M, and III). Here, employing the scheme developed in Reference 37, we calculate, in each region, the bulk complex band structure $E(k_\tau^{(N)}; k_x)$, where $k_\tau^{(N)}$ is the y-component of bulk wave vector relative to the corresponding Dirac point with valley index $\tau$, in the region. In general, $k_\tau^{(N)}$ is a complex number. For a given E and a given $k_x$, there are four bulk solutions for each valley, with $k_\tau^{(N)}$'s and the corresponding wave functions being given, respectively, by $k_{\tau,j}^{(N)}$ and $e^{ik_x x}e^{ik_{\tau,j}^{(N)}y}\Psi_{\tau,j}^{(N)}$, for j = 1-4. $\Psi_{\tau,j}^{(N)}$ here denotes a four-component column vector. Next, we form a general bulk solution in each region, and write

$$\Psi^{(N)}(\vec{r}) = e^{i\vec{K}\cdot\vec{r}}\Psi_+^{(N)}(\vec{r}) + e^{-i\vec{K}\cdot\vec{r}}\Psi_-^{(N)}(\vec{r}),$$
$$\Psi_\tau^{(N)}(\vec{r}) = e^{ik_x x}\sum_{j=1}^{4}c_{\tau,j}^{(N)}e^{ik_{\tau,j}^{(N)}y}\Psi_{\tau,j}^{(N)}. \quad (6)$$

We match the bulk solutions between different regions in such a way that cell-averaged current continuity at the interface is satisfied [50], and also enforce the hard-wall boundary conditions at defect lines realized by, e.g., the edges of a nanoribbon or graphene oxide [51]: $\Psi^{(I)}(\vec{r}) = 0$ at $y = W_1 + W_2/2$ and $\Psi^{(III)}(\vec{r}) = 0$ at $y = -W_2/2 - W_3$ at $y = W_1 + W_2/2$ and $y = -W_2/2 - W_3$. This determines, for a given $k_x$, the corresponding subband energy levels, and hence $E(k_x; n)$, the subband structure (n = subband index).

## IV. RESULTS

We consider the BLG-based Type-I valley filter configuration specified by the parameters taken from **Figure 5(d)**, namely, $W_1 = W_3 = 90$ Å, $W_2 = 100$ Å, $\Delta_1 = \Delta_3 = 116$ meV, $\Delta_2 = 17$ meV, and conduction band offset = 99 meV. For the in-plane potential energy difference, we take $V_1 - V_3 = 70$ meV. **Figure 8(a)** shows the subband dispersion for the corresponding quantum wire with $W_1 = W_3 = \infty$. In this case the scattering defects are located far

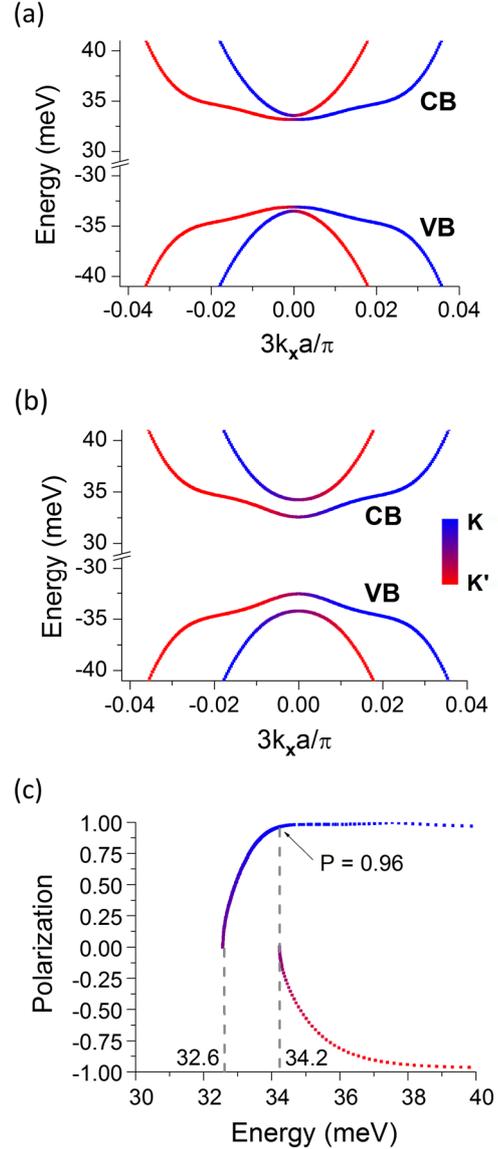

**Figure 8.** Subbands and polarization for the BLG-based Type-I filter configuration with $W_1 = W_3 = 90$ Å, $W_2 = 100$ Å, $\Delta_1 = \Delta_3 = 116$ meV, $\Delta_2 = 17$ meV, $V_1 - V_3 = 70$ meV, and conduction band offset = 99 meV. **(a)** Conduction bands (CB) and valence bands (VB) in the quantum wire calculated by taking $W_1 = W_3 = \infty$. Blue curves: subbands of K pseudospin.

Red curves: subbands of K' pseudospin. **(b)** Subbands in the quantum wire with scattering defects taken into account, showing the opening of a pseudogap (~ 1.6 meV). **(c)** Valley polarization for right-moving electron states. Blue solid line: for states of the first conduction subband inside the pseudogap; blue dotted line: for states of the first conduction subband above the pseudogap; and red dotted line: for states of the second conduction subband.

away from the quantum wire and do not have any effect on the subband dispersion. Due to the difference between $V_1$ and $V_3$, a nonvanishing in-plane electric field is present and breaks the valley degeneracy leading to the Rashba pseudospin splitting, as shown in the figure. With the parameters given here, the size of energy splitting at $k_x \sim 0.01\pi/3a$ can reach about 2.2 meV. **Figure 8(b)** shows the subbands with the effect of scattering defects taken into account. A pseudogap of about 1.6 meV in size is opened due to the scattering-induced coupling between the two states of opposite pseudospins.

Next, based on the calculation of subband structure, we discuss, in **Figure 8(c)**, the valley polarization of an electron passing through the specific valley filter. We define the corresponding valley polarization by

$$P \equiv \frac{P_+ - P_-}{P_+ + P_-}, \quad (7)$$
$$P_+ \equiv \int d^2r |\Psi_+(\vec{r})|^2, \quad P_- \equiv \int d^2r |\Psi_-(\vec{r})|^2.$$

where $\Psi_+(\vec{r})$ and $\Psi_-(\vec{r})$ are the two pseudospin components in Eqn. (5). **Figure 8(c)** shows the result of valley polarization vs. electron energy, for right-moving electron states. The polarization vanishes for electron energy near the bottom of the pseudogap (at 32.6 meV), increases rapidly, and reaches the maximum near the top of the pseudogap (at 34.2 meV). For the specific structure considered here, the optimal polarization can reach almost unity. We note that this optimal filtering performance can be achieved by placing in front of the valley filter an additional energy filter (for example, a resonant tunneling structure) that passes only electrons with energy near the top of the pseudogap. The combination of an energy filter and a valley filter provides a set-up for room temperature valley filtering as well, when the energetically filtered electrons are sent right into the valley filter before any energy relaxation process occurs.

Before presenting further numerical results of polarization, we note below two useful symmetry-based properties of the polarization. Specifically, we focus on the symmetric valley filter defined by $W_1 = W_3$, $\Delta(y) = \Delta(-y)$, and $V(y) = -V(y)$. Here, V and Δ do not have to be piecewise constant.

### Effect of reversing the in-plane electric field

We consider an electron state in the symmetric structure, with the wave function for a specific $k_x$ being given by $\Psi(\vec{r}) = e^{i\vec{K}\cdot\vec{r}}\Psi_+(\vec{r}) + e^{-i\vec{K}\cdot\vec{r}}\Psi_-(\vec{r})$, where

$$\Psi_+(\vec{r}) = e^{ik_x x}(A_1(y), \ B_1(y), \ A_2(y), \ B_2(y))^t$$
$$\Psi_-(\vec{r}) = e^{ik_x x}(A_1'(y), \ B_1'(y), \ A_2'(y), \ B_2'(y))^t$$
(8)

The effect of reversing the in-plane electric field can be investigated by considering a potential $U(y)$ with $U(y) = -V(y)$. It can be verified that, with the reversal, a solution, denoted by $\Phi(\vec{r})$ and degenerate with $\Psi(\vec{r})$ (meaning that both have the same wave vector and the same energy) exists and is given by $\Phi(\vec{r}) = e^{i\vec{K}\cdot\vec{r}}\Phi_+(\vec{r}) + e^{-i\vec{K}\cdot\vec{r}}\Phi_-(\vec{r})$, where

$$\Phi_+(x, y) = \Psi_-(x, -y) \quad (9)$$
$$\Phi_-(x, y) = \Psi_+(x, -y)$$

Notice the switch between K and K' amplitudes in $\Phi(\vec{r})$, when compared to those in $\Psi(\vec{r})$. In other words, if the original electron state is primarily K-polarized (i.e., $\|\Psi_+(\vec{r})\| \gg \|\Psi_-(\vec{r})\|$) with valley polarization $P$, then reversing the field will lead to K'-polarization with the polarization given by $-P$. This permits us to electrically switch the device between K and K' valley polarization.

### Effect of reversing $k_x$

One can easily verify that in association with the solution $\Psi(\vec{r})$, there is always a solution $\Theta(\vec{r}) = e^{i\vec{K}\cdot\vec{r}}\Theta_+(\vec{r}) + e^{-i\vec{K}\cdot\vec{r}}\Theta_-(\vec{r})$, where

$$\Theta_+(\vec{r}) = \Psi_-(\vec{r})^* = e^{-ik_x x}(A_1'(y)^*, \ B_1'(y)^*, \ A_2'(y)^*, \ B_2'(y)^*)^t$$
$$\Theta_-(\vec{r}) = \Psi_+(\vec{r})^* = e^{-ik_x x}(A_1(y)^*, \ B_1(y)^*, \ A_2(y)^*, \ B_2(y)^*)^t$$
(10)

which is degenerate with $\Psi(\vec{r})$ in energy but with the wave vector being reversed, e.g., $k_x \to -k_x$. As expressed in Eqn. (10), the amplitudes of the valleys are switched here in comparison to those in $\psi(\vec{r})$. Therefore, if the state with $k_x$ carries polarization $P$, then the state with $-k_x$ carries the reversed polarization $-P$.

The structural parameters can be varied to increase the operational energy range of the valley filter, as shown in **Figure 9**. The example considered is specified by the parameters taken from **Figure 6(d)**, namely, $W_1 = W_3 = 40$ Å, $W_2 = 50$ Å, $\Delta_1 = \Delta_3 = 149$ meV, $\Delta_2 = 127$ meV, and conduction band offset = 125 meV. For the in-plane potential energy difference, we take $V_1 - V_3 = 80$ meV. **Figure 9** demonstrates valley filtering in a Type-II structure. It also shows that a large pseudogap (~ 16.7 meV) can be achieved with the optimal polarization simultaneously reaching about 66%. When the size of pseudogap is tuned around the room temperature thermal energy, it is speculated that the filter alone may be operated near room temperature by placing the Fermi energy inside the pseudogap.

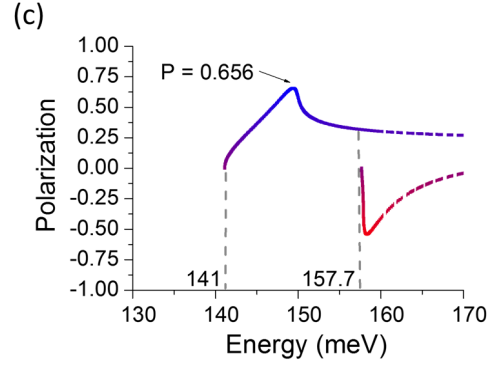

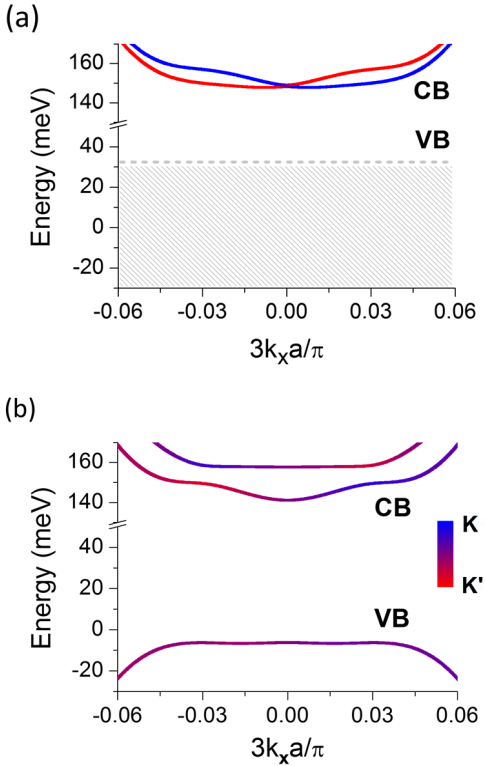

**Figure 9.** Subbands and polarization for the BLG-based Type-II filter configuration with $W_1 = W_3 = 40$ Å, $W_2 = 50$ Å, $\Delta_1 = \Delta_3 = 149$ meV, $\Delta_2 = 127$ meV, $V_1 - V_3 = 80$ meV, and conduction band offset = 125 meV. **(a)** Conduction bands (CB) and valence bands (VB) in the quantum wire calculated by taking $W_1 = W_3 = \infty$. Blue curves: subbands of K pseudospin. Red curves: subbands of K' pseudospin. Continuum region: extended valence band states outside the channel. **(b)** Subbands in the quantum wire with scattering defects taken into account, showing the opening of a pseudogap (~ 18 meV). **(c)** Valley polarization for right-moving electron states. Blue solid line: for states of the first conduction subband inside the pseudogap; blue dotted line: for states of the first conduction subband above the pseudogap; and red dotted line: for states of the second conduction subband.

In **Figure 10**, we consider the SLG-based Type-I valley filter configuration specified by $W_1 = W_3 = 170$ Å, $W_2 = 3050$ Å, $\Delta_1 = \Delta_3 = 15$ meV, $\Delta_2 = 0$ meV, $V_1 - V_3 = 10$ meV, and conduction band offset = 15 meV. It shows the opening of a pseudogap ~ 0.8 meV with the optimal polarization ~ 70%. We note that the small pseudogap size is adjustable for potential room temperature applications. For example, it may be increased by reducing the well width $W_2$. This would move up the energy subband with a deeper penetration of electrons into the barrier for a stronger ribbon edge scattering-induced intervalley mixing to expand the pseudogap. In doing so, on the other hand, the resultant valley polarization would be reduced by the stronger intervalley mixing.

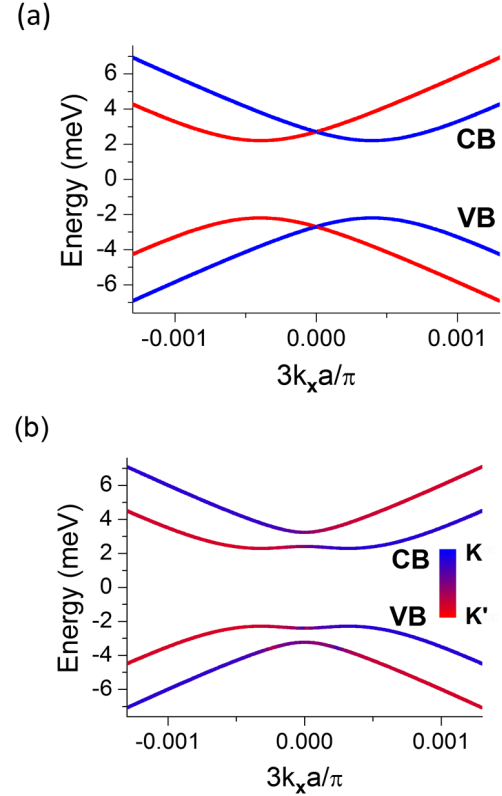

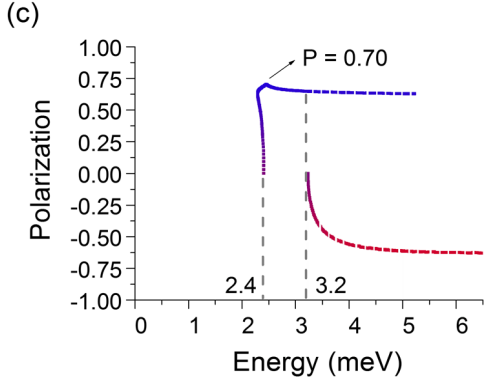

**Figure 10.** Subbands and polarization for the SLG-based Type-I filter configuration with parameters specified by $W_1 = W_3 = 170$ Å, $W_2 = 3050$ Å, $\Delta_1 = \Delta_3 = 15$ meV, $\Delta_2 = 0$ meV, $V_1 - V_3 = 10$ meV, and conduction band offset = 15 meV. **(a)** Conduction bands (CB) and valence bands (VB) in the quantum wire calculated by taking $W_1 = W_3 = \infty$. Blue curves: subbands of K pseudospin. Red curves: subbands of K' pseudospin. **(b)** Subbands in the quantum wire with scattering defects taken into account, showing the opening of a pseudogap (~ 0.8 meV). **(c)** Valley polarization for right-moving electron states. Blue solid line: for states of the first conduction subband inside the pseudogap; blue dotted line: for states of the first conduction subband above the pseudogap; and red dotted line: for states of the second conduction subband.

## V. SUMMARY

In summary, an all-electrical quantum wire configuration made of gapped graphene has been proposed that can function as a valley filter and a half valley valve for building a valleytronic circuit. Such a structure is featured by i) consistency with planar processing, ii) potential for room temperature applications, and iii) robustness against configuration fluctuation. Moreover, due to the electrical controllability of its valley polarity, the proposed structure can be utilized as the interface between a valleytronic circuit and an electronic circuit, therefore opening up an interesting path to integrated electro-valleytronics.

In **W-II**, systematic numerical calculations will be performed to study electron transport through valley valves. Using valves formed by the VOI-based filters proposed here as examples, we are able to investigate effects specific to the VOI-based valves, as well as those in generic, practical valves such as effects due to impurity scattering, inter-filter spacing, the increasing number of filters, and etc.

**Acknowledgment** – We would like to thank Yen-Chun Chen, Jia-Huei Jiang and Yen-Ju Lin for their assistance in the numerical simulation of electrostatic potential distribution in the proposed structure. We would also like to acknowledge the financial support of MoST, ROC through Contract No. MOST 109-2811-M-007-561 and the Thematic Project of Academia Sinica.